\documentclass[sort
    ,final            % use final for the camera ready runs
  ]
  {aipproc}

\layoutstyle{8x11double}

\newcommand\psr{1E~1048.1$-$5937}
\newcommand\xte{\textit{RXTE}}
\newcommand\cxo{\textit{CXO}}
\newcommand\xmm{\textit{XMM}}
\newcommand\swift{\textit{Swift}}
\newcommand\hst{\textit{HST}}
\newcommand\vlt{\textit{VLT}}

\begin{document}

\title[The Variable X-ray and Near-IR Behaviour of \psr]{The
  Variable X-ray and Near-IR Behavior of the Particularly Anomalous
  X-ray Pulsar \psr}


\author{Cindy R. Tam}{
  address={Department of Physics, McGill University, 3600 University
  Street, Montreal, QC, H3A 2T8, Canada}
}

\author{Fotis P. Gavriil}{
  address={NASA Goddard Space Flight Center, Astrophysics Science
  Division, Code 662, Greenbelt, MD, 20771, USA}
}

\author{Rim Dib}{
  address={Department of Physics, McGill University, 3600 University
  Street, Montreal, QC, H3A 2T8, Canada}
}

\author{Victoria M. Kaspi}{
  address={Department of Physics, McGill University, 3600 University
  Street, Montreal, QC, H3A 2T8, Canada}
}

\author{Peter M. Woods}{
  address={Dynetics, Inc., 1000 Explorer Boulevard, Huntsville, AL,
  35806, USA}
  ,altaddress={NSSTC, 320 Sparkman Drive, Huntsville, AL, 35805, USA} % additional visiting address
}

\author{Cees Bassa}{
  address={Department of Physics, McGill University, 3600 University
  Street, Montreal, QC, H3A 2T8, Canada}
}

\begin{abstract}
We present the results of X-ray and near-IR observations of the
anomalous X-ray pulsar \psr, believed to be a magnetar.
This AXP underwent a period of extreme variability during 2001-2004,
but subsequently entered an extended and unexpected quiescence in
2004-2006, during which we monitored it with \xte, \cxo, and \hst.
Its timing properties were stable for $>$3 years throughout the
quiescent period. \psr\ again went into outburst in March 2007, which
saw a factor of $>$7 total X-ray flux increase which was
anti-correlated with a pulsed fraction decrease, and correlated with
spectral hardening, among other effects. The near-IR counterpart also
brightened following the 2007 event. We discuss our findings in the
context of the magnetar and other models.
\end{abstract}

%%%%%%%%%%%%%%%%%%%%%%%%%%%%%%%%%%%%%%%%%%%%%%%%%%%%%%%%%%%%%%%%%%%
%%
%% The below \maketitle command inserts the actual front matter data.
%% It has to follow the above declarations.
%%
%%%%%%%%%%%%%%%%%%%%%%%%%%%

\maketitle

%%%%%%%%%%%%%%%%%%%%%%%%%%%%%%%%%%%%%%%%%%%%
%% MAINMATTER
%%
%%%%%%%%%%%%%%%%%%%%%%%%%%%%%%%%%%%%%%%%%%%%%%%%%%%%%%%%%%%%%%%%%%%%%%%%%%%%
%% Headings:
%%
%% The aipproc supports three heading levels, i.e., \section,
%%	\subsection, and \subsubsection.
%%%%%%%%%%%%%%%%%%%%%%%%%%%%%%%%%%%%%%%%%%%%%%%%%%%%%%%%%%%%%%%%%%%%%%%%%%%%
%% Cross-references:
%%
%% Page numbers (\pageref) and headings can NOT be referenced in the class,
%% since before being produced, no page numbers are determined.
%%
%% Tables, figures, and equeations can be referenced by using the LaTex
%% 	commands \label and \ref. For references to equation numbers, \eqref
%%	can be used, which will print "(1)" (while \ref will result in "1").
%%
%%%%%%%%%%%%%%%%%%%%%%%%%%%%%%%%%%%%%%%%%%%%%%%%%%%%%%%%%%%%%%%%%%%%%%%%%%%%
%% Lists: 
%%
%% Standard "itemize", "enumerate", etc. list environments are supported.
%%%%%%%%%%%%%%%%%%%%%%%%%%%%%%%%%%%%%%%%%%%%%%%%%%%%%%%%%%%%%%%%%%%%%%%%%%%%
%% Urls:
%%
%% \url{} command is provided for documenting URLs.
%%%%%%%%%%%%%%%%%%%%%%%%%%%%%%%%%%%%%%%%%%%%

\section{Introduction}

Anomalous X-ray pulsars\footnote{For a summary of observed properties,
see \url{http://www.physics.mcgill.ca/~pulsar/magnetar/main.html}.}
(AXPs) are a class of pulsars with observed X-ray luminosities in
excess of what can be provided by rotational spin-down.  
Observationally, they possess extremely large inferred surface
magnetic fields ($\sim10^{14}$~G), exhibit a variety
of variable behaviour such as X-ray bursts, timing glitches and
changes to their flux and spectrum, and are now known to be emitters
of optical and infrared (IR), and in some cases even radio radiation. 
It is believed that they, like the soft gamma repeaters (SGRs), are
magnetars: isolated neutron stars powered by the decay of enormous
magnetic fields \citep{td95,td96a}.  The resulting
effects of the magnetic field on the crust and magnetosphere produce
the observational signatures found in AXPs and SGRs.

Monitoring of AXP timing properties with the \textit{Rossi X-ray
Timing Explorer} has been ongoing for $>$10 years.  The 6.45-s pulsar
\psr\ is an AXP with a particularly unique history. 
Prior to 2004, this pulsar was highly rotationally unstable, so that
phase coherence could only been maintained for months at a
time \citep{gk04}. During 2001-2002, \psr\ underwent two prolonged
flux ``flares'' (not to be confused with SGR giant flares) that
were unlike behaviour seen previously in any other magnetar. The 
time-resolved flux increases took place over $\sim$weeks, and the
gradual decay of the larger flare lasted years
\citep{mts+04,gk04,gkw06}. At the same time, erratic torque
variability, X-ray spectral variablity \citep{tmt+05}, SGR-like
bursts, and a near-IR flux enhancement \citep{wc02,ics+02} occurred.
Thus, we proposed for simultaneous monitoring observations with the 
\textit{Chandra X-ray Observatory} and the \textit{Hubble Space
Telescope} in 2006.  

\section{Pre-2007 ``Quiescence''}

\begin{figure}
  \includegraphics[width=0.43\textwidth]{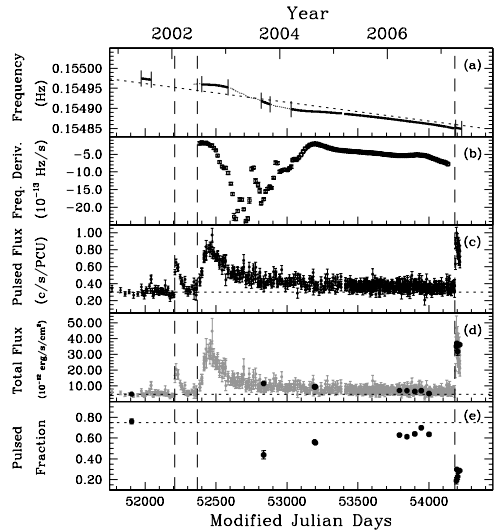}
  \caption{The evolution of \psr's rotational and pulsed
  properties; figure from \citet{tgd+07}. Fluxes and pulsed fraction
  are given for 2$-$10~keV.
  (\textit{a}) Spin frequency as measured with \xte\ monitoring.
  (\textit{b}) Frequency derivative; see also \citet{gk04}.
  (\textit{c}) \xte-derived pulsed flux.
  (\textit{d}) Simulated total unabsorbed flux, described in the
  text.
%  (\textit{d}) Total unabsorbed flux as measured
%  with \cxo, \swift, and \xmm\ (\textit{white}).  Also shown are
%  simulated total flux data (\textit{black}), described in the text.
  (\textit{e}) Pulsed fraction.
  \label{fig:timing}}
\end{figure}

\begin{figure}
  \includegraphics[width=.45\textwidth]{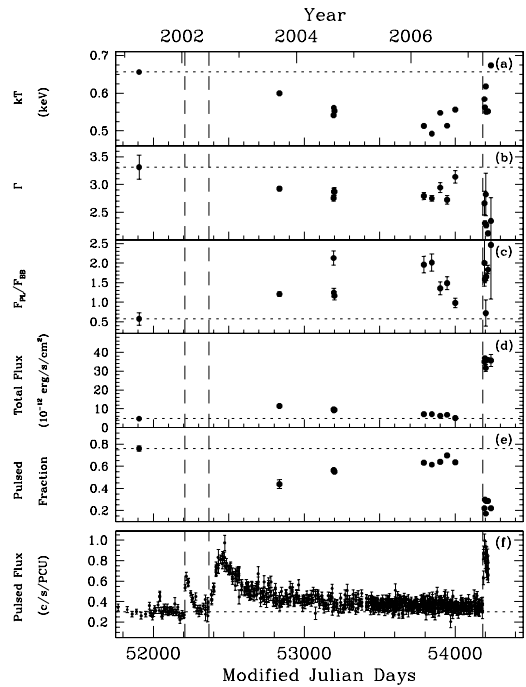}
  \caption{The X-ray spectral history of \psr\ \citep{tgd+07}. All
  spectral data from \cxo, \swift, and \xmm\ were jointly fit
  to an absorbed blackbody plus power law model, with a single
  best-fit value of $N_H = (0.97\pm0.01)\times
  10^{22}$~cm$^{-2}$. Fluxes and pulsed fractions are given for
  2$-$10~keV.
  (\textit{a}) Blackbody temperature.
  (\textit{b}) Photon index.
  (\textit{c}) Ratio of blackbody to power-law flux
  contributions.
  (\textit{d}) Total unabsorbed flux.
  (\textit{e}) Pulsed fraction.
  (\textit{f}) \xte-derived pulsed flux.
  \label{fig:xray}}
\end{figure}

Between 2004-2006, \psr\ appeared to be in a state of relative
quiescence. Our \xte\ observations showed that the pulsed flux was at
the same level as during its pre-2001 quiescence, and a phase coherent
timing solution was maintained for $>$3 years, longer than ever before
maintained; see Figure~\ref{fig:timing}\textit{a}.  \cxo\
observations in 2006 revealed that the X-ray spectrum, when fit to an
absorbed blackbody plus power-law model, had varied slightly, but
intruigingly, did not return to quiescent levels like the pulsed and
total flux did; see Figure~\ref{fig:xray}.  In 
February 2006, \psr\ was detected using the \hst\ filter F160W
(similar to $H$-band) at a lower flux level than all previous detections
($m_{F160W}=22.70\pm0.14$~mag); in subsequent observations, it had
dropped below detectability (see 
Fig.~\ref{fig:nir}\textit{b}).  We also analysed
archival \textit{Very Large Telescope} observations, and detected the
counterpart once faintly at $K_S=21.0\pm0.3$~mag in 2005
(Fig.~\ref{fig:nir}\textit{c}). These results are presented in more
detail in \citet{tgd+07}.

\section{The March 2007 Event}

In late March 2007, the quiescent phase unexpectedly ended
with the sudden reactivation of \psr\ in a new flare, discovered
through our \xte\ monitoring (Dib et al. ATel \#1041). The details of
our analysis were originally published in \citet{tgd+07}.

\paragraph{X-ray results} 
Simultaneous with a large glitch, \xte\ saw a factor of $\sim$3
increase in the pulsed flux (2$-$10~keV), with a rise time of
$<$1 week (Fig.~\ref{fig:timing}\textit{c}); details of the \xte\
results will be presented by Dib et al. (in preparation).  Follow-up
observations with \cxo\ (Gavriil et al. ATel \#1076), \swift\ (Campana
et al. ATel \#1043, Israel et al. ATel \#1077), and
\textit{XMM-Newton} (Rea et al. ATel \#1121) revealed that the total
flux initially increased by a factor of $>$7 (2$-$10~keV) relative to
the quiescent flux, while the pulsed fraction decreased from
$\sim$75\% to $\sim$20\%.  We also observed a spectral hardening
(Fig.~\ref{fig:xray}\textit{b}) correlated with the flux increase, and
a change in the pulse profile from nearly sinusoidal to having
multiple peaks after the flare \citep{tgd+07}.

\begin{figure}
  \includegraphics[width=.3\textwidth]{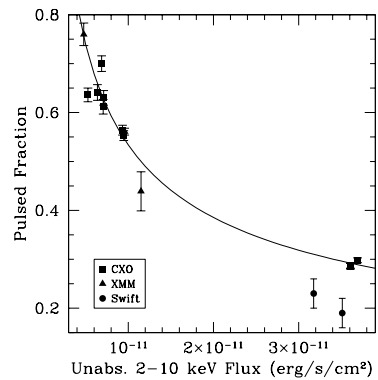}
  \caption{Pulsed fraction vs. total unabsorbed flux, from \cxo,
  \swift, and \xmm\ measurements in 2$-$10~keV \citep{tgd+07}. The
  curve indicates the best-fit power law that describes the
  correlation. \label{fig:pulsefrac}}
\end{figure}

We confirmed the anti-correlation between total X-ray flux and pulsed
fraction (Figs.~\ref{fig:xray}\textit{d}-\ref{fig:xray}\textit{e})
noted previously by \citet{tmt+05} and \citet{gkw06}. The clear
dependence, shown in Figure~\ref{fig:pulsefrac}, is well described by
a power law \citep{tgd+07}.  Given this
relationship, and the definition of pulsed fraction as pulsed flux
divided by total flux, we can simulate a well sampled data set,
demonstrating how we \emph{expect} the total flux behaved in the
past.  This is shown in Figure~\ref{fig:timing}\textit{d}.

\paragraph{AXP emission models}
The source of the anti-correlation between pulsed fraction and total
flux is not obvious.  In principle, it could be the result
of a growing hot spot on the magnetar surface, produced by either
changing internal processes, or changes in returning magnetospheric
currents.  This is complicated by such complex effects as surface
thermal emission \citep{og07}, light bending and radiative beaming,
and magnetospheric scattering \citep{tlk02,ft07}.  The observed
hardness-intensity correlation is predicted by the twisted magnetic
field model \citep{tlk02}, and can also possibly be explained by
surface thermal emission \citep{og07}.

An alternative model for AXP emission is accretion from a fallback
debris disk around an ordinary pulsar \citep{chn00}, in which both the
pulsar's spin-down rate (Fig.~\ref{fig:timing}\textit{b}) and X-ray
luminosity (Fig.~\ref{fig:timing}\textit{d}) are heavily dependent on
the mass accretion rate.  \citet{gk04} showed that for a pulsar
undergoing spin down, we would expect $L_X \propto |\dot{\nu}|^{7/3}$.
However, we find that the factor of $>$10 variability in $\dot{\nu}$
between 2002-2004 was not reflected in a factor of $>$200 change in
X-ray luminosity as expected; in fact, the unabsorbed flux changed by
merely $\sim$6$\times$, and asynchronously with $\dot{\nu}$.

\begin{figure}
  \includegraphics[width=.45\textwidth]{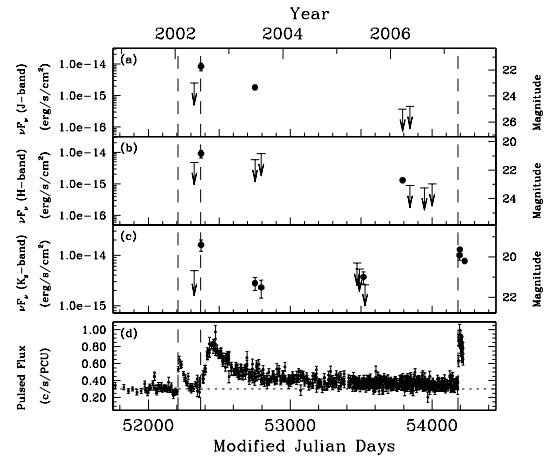}
  \caption{The near-IR flux history of \psr. \vlt\ and \hst\
  measurements from 2005-2006 are from this analysis \citep{tgd+07};
  data prior to that are from previous literature
  \citep{wc02,ics+02,dv05a}. Show on the right axis are the
  approximate $JHK_S$ magnitudes.
  \label{fig:nir}}
\end{figure}

\paragraph{Near-IR results}
Following the March 2007 event, optical/IR observations were obtained
with the \textit{Magellan Telescope} (Wang et al. ATel \#1044) and the
\vlt\ (Wang et al. ATel \#1071, Israel et al. ATel \#1077); see
Figure~\ref{fig:nir}.  A detailed analysis will be forthcoming (Wang
et al. in preparation). These new observations 
showed that the optical/IR flux may in fact be correlated with X-ray
flux, contrary to what was previously suggested \citep{dv05a}.
Near-IR variability is seen as correlated with X-rays in another AXP
\citep{tkvd04}, but appears uncorrelated in two other cases
\citep{dv06c,crp+07}.  Such inconsistent behaviour is puzzling for the
accretion disk model, since near-IR radiation is 
thought to be closely tied to X-ray emission via reprocessing of the
X-rays in the disk. Optical/IR emission from magnetars
has been attributed to high-energy processes (such as curvature
or ion cyclotron emission) occurring in the magnetosphere
\citep{bt07}. Regardless, more frequent monitoring of AXP variability
will be required in order to set constraints on optical/IR models.

\section{Conclusions}

From $\sim$10 yrs of multiwavelenth observations, it is apparent that
all spin and radiative activity from \psr\ prior to 2004 can be linked
to the large flares of 2001-2002, and that this AXP was in a relative
quiescence between 2004-2006, which ended in March 2007 with another
flaring event.  We speculate that the observed behaviour, such as the
varying X-ray flux anti-correlated with pulsed fraction and correlated
with hardness, may be consistent with current magnetar scenarios.

\end{document}